\documentclass[12pt]{article}
\usepackage{epsfig}
\usepackage{amsmath,graphicx,axodraw}

\def\Li#1#2{{\mathrm{Li}}_{#1}\left(#2\right)}

\def\ba{\begin{eqnarray}}
\def\ea{\end{eqnarray}}

\newcommand{\FF}{\mathcal{F}}
\def\dd{{\mathrm d}}

\def\fun#1#2{\lower3.6pt\vbox{\baselineskip0pt\lineskip.9pt
  \ialign{$\mathsurround=0pt#1\hfil##\hfil$\crcr#2\crcr\sim\crcr}}}
\def\order#1{{\mathcal O}\left(#1\right)}
\newcommand{\vecc}[1]{\mbox{\boldmath $#1$}}
\def\gsim{\mathrel{\raise.3ex\hbox{$>$\kern-.75em\lower1ex\hbox{$\sim$}}}}
\def\lsim{\mathrel{\raise.3ex\hbox{$<$\kern-.75em\lower1ex\hbox{$\sim$}}}}
\newcommand{\GeV}{\unskip\,\mathrm{GeV}}

\title{Radiative Corrections to High Energy Lepton Bremsstrahlung on Heavy Nuclei
}

\author{Andrej B. Arbuzov \\[.2cm]
{\it Joint Institute for Nuclear Research, 141980 Dubna, Russia} \\
{\tt E-mail: arbuzov@theor.jinr.ru}
}

\date{}

\begin{document}

\maketitle

\begin{abstract}
One-loop radiative corrections to the leptonic tensor in high energy
bremsstrahlung on heavy nuclei are calculated. Virtual and real photon
radiation is taken into account. Double bremsstrahlung is simulated
by means of Monte Carlo. Numerical results are presented for the case of 
muon bremsstrahlung in conditions of the COMPASS experiment at CERN.
\end{abstract}

\section{Introduction}
\label{sect1}

Charged lepton Bremsstrahlung on nuclei has been studied both theoretically
and experimentally for many years (see {\it e.g.} 
textbooks~\cite{Akhiezer1,BLP} and references therein). This process 
contributes to energy losses of a lepton propagation through matter, which
is relevant for many applications.

The similar process of high energy pion bremsstrahlung is used for 
extraction of pion polarizability~\cite{Antipov:1982kz,Moinester:2000eq}.
In the modern COMPASS experiment~\cite{Abbon:2007pq,COMP_prop} 
the muon bremsstrahlung is used as a reference cross section and for estimates 
of systematic uncertainties. 
For this reason differential distribution of the muon bremsstrahlung should 
be predicted with high accuracy. That requires to take
into account several effects beyond the tree-level Born approximation.
So one needs to consider  multiple photon exchange with the nucleus 
(important for large $Z$ values), electromagnetic nuclear elastic 
and inelastic form factors, screening of the nucleus by the electrons 
surrounding it, and inelastic interactions of the projectile particle with the
atomic electrons (see Ref.~\cite{Andreev:1997pf} and references therein). 
Besides those, we have to take into account also 
the vacuum polarization in the exchanged photon and 
at least one-loop radiative corrections to the lepton tensor.
In this paper a new calculation of the latter is presented.

Ref.~\cite{Vanderhaeghen:2000ws} gives a comprehensive report on the calculation of one-loop
corrections to virtual Compton scattering $(ep\to ep\gamma)$. The lepton bremsstrahlung on
a heavy nucleus we met here is a specific case of the general problem.
Results of Ref.~\cite{Vanderhaeghen:2000ws} for the Compton tensor also can't 
be directly applied to the
problem under consideration, since the actual kinematical 
conditions~(see Eq.(\ref{kinem_cond}) below) deserve a special treatment.
The latter includes keeping an exact dependence on the lepton mass and
providing numerical stability of the corresponding computer code. 

Since for this kinematical region we have to keep the exact dependence
on the lepton mass, analytic formulae for the corrections become lengthy and
cumbersome. Moreover to provide a possibility to impose various experimental
cuts, we perform integration over the final state phase space numerically.

The paper is organized as follows. In the next section we give the notation 
and the explicit expression for the Born cross section. Sect.~\ref{Sect_Corr}
presents the calculation of various radiative correction contributions: the one due
to a single virtual loop, the one due to additional soft photon emission, and the
one due to double bremsstrahlung. Numerical results and conclusions 
are given in the last section.

\section{Preliminaries}
\label{Sect_Prel}

At the Born level we can represent the differential spectrum of the hard photon 
produced in the process
\ba
l(p_1)\ +\ A(P)\ \to\ l(p_2)\ +\ \gamma(k) + A(P')
\ea
in the form~\cite{Akhiezer1,BLP}
\ba\label{born_a}
\frac{\dd\sigma^{\mathrm{Born}}}{\dd\omega} &=& 
 \frac{Z^2\alpha^3}{2\pi}\int\limits_{-1}^{1}\dd c_1
\int\limits_{-1}^{1}\dd c_2\int\limits_{0}^{2\pi}\dd \varphi\,
\frac{|\vecc{p}_2|}{|\vecc{p}_1|}\frac{\omega}{Q^4}
    \biggl( \frac{|\vecc{p}_2|^2}{\chi_2^2} (4E_1^2-Q^2)s_2^2
 + \frac{|\vecc{p}_1|^2}{\chi_1^2}(4E_2^2-Q^2)s_1^2 
\nonumber \\
&+& 2\frac{\omega^2}{\chi_1\chi_2}(|\vecc{p}_1|^2s_1^2+|\vecc{p}_2|^2s_2^2)
 - 2\frac{|\vecc{p}_1| |\vecc{p}_2|}{\chi_1\chi_2}(2E_1^2+2E_2^2-Q^2)
  s_1s_2\cos{\varphi}
 \biggr),
\ea
where $\omega=k^0$ is the emitted photon energy; $\varphi$ is the azimuthal
angle of the scattered lepton; $Z$ is the nucleus charge; $m$ is the lepton mass;
$E_{1(2)}$ and $p_{1(2)}$ are the energies and 4-momenta of the projectile (scattered)
leptons,
\ba
c_{1,2} = \cos(\widehat{\vecc{k}\vecc{p}}_{1,2}), \quad 
s_{1,2} = \sin(\widehat{\vecc{k}\vecc{p}}_{1,2}), \quad 
\chi_{1,2} = kp_{1,2}, \quad
Q^2 = - (p_1 - p_2 - k)^2.
\ea
Here and in what follows, it is assumed that the lepton mass is small compared
with the atom mass, while the energies are large:
\ba
m \ll M_A,\qquad E_1 \gg m,\qquad E_2 \gg m,\qquad \omega \gg m.
\ea

Let us rewrite the Born cross section~(\ref{born_a}) via a set of form factors:
\ba
\frac{\dd\sigma^{\mathrm{Born}}}{\dd\omega} &=& 
 \frac{Z^2\alpha^3}{2\pi}\int\limits_{-1}^{1}\dd c_1
\int\limits_{-1}^{1}\dd c_2\int\limits_{0}^{2\pi}\dd \varphi\,
\frac{|\vecc{p}_2|}{|\vecc{p}_1|}\frac{\omega}{Q^4}
    \frac{1}{2e^4}\biggl( \FF^{(0)}_{\delta}(q_s,t_s,u_s)
- \FF^{(0)}_{11}(q_s,t_s,u_s)E_2^2
\nonumber \\
&-& \FF^{(0)}_{22}(q_s,t_s,u_s)E_1^2
+ \FF^{(0)}_{12}(q_s,t_s,u_s)E_2E_1
+ \FF^{(0)}_{21}(q_s,t_s,u_s)E_1E_2
  \biggr),
\\ \nonumber 
q_s&=& 2\chi_2 - 2\chi_1 +Q^2,\qquad  
t_s=2\chi_1-m^2,\qquad u_s = -2\chi_2-m^2,
\ea
where $e$ is the electron charge.
The notation in the above expression is adjusted to the one
used in the SANC~\cite{Andonov:2004hi} system, where the relevant
expressions can be found as for the Born-level form factors
as well as for the ones in the one-loop approximation.

Studying the differential distribution in the scattering angles of
the Born cross section, one can see that it is peaked in the kinematical
domain, where
\ba \label{kinem_cond}
\widehat{\vecc{k}\vecc{p}}_{1,2} \sim \widehat{\vecc{p}_1\vecc{p}}_2 \sim \frac{m}{E_1}. 
\ea
For the case of high energy muon scattering $(E_1\sim 100$~GeV) being under consideration now,
the angles become small. Moreover, one should be careful with the dependence on the lepton
mass, since $m^2 \sim \chi_{1,2}$ in this domain. On the other hand, we can safely
drop some terms, proportional to the small ratio $m^2/E_1^2$. As concerning the momentum
transferred, contrary to the case of the Rutherford scattering, it can't go down below the
kinematical threshold value
\ba
Q_{\mathrm{min}} \equiv \sqrt{Q^2_{\mathrm{min}}} = \frac{m^2\omega}{2E_1E_2}\, .
\ea 
In the ultra-relativistic approximation applicable in our case, after an integration over 
the whole phase space, one gets the Born-level photon spectrum in the simple form
\ba
\frac{\dd\sigma}{\dd\omega} &=& \frac{4Z^2\alpha^3}{m_{\mu}^2\omega E_1^2}
\biggl(E_1^2 + E_2^2 - \frac{2}{3}E_1E_2\biggr), \qquad \omega = E_1 - E_2.
\ea

\section{One-Loop Corrections}
\label{Sect_Corr}

We subdivide the contributions of the one-loop QED corrections into three parts:
1) the one due to a single virtual loop; 2) the one due to soft real photon 
emission; 3) and the one due to additional hard photon emission (double bremsstrahlung).

\subsection{Virtual Loop Contribution}
\label{subSect_V}

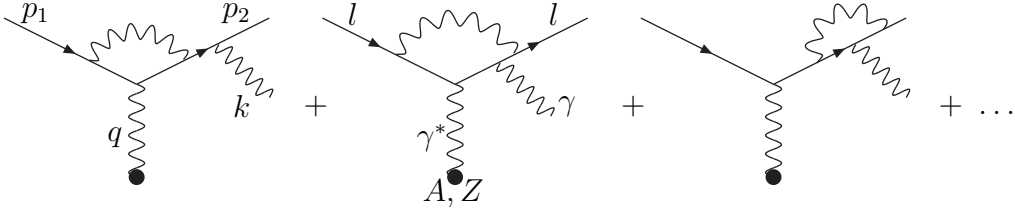
\begin{figure}[htb]
\begin{picture}(400,80)(0,0)
\ArrowLine(10,70)(60,45)
\ArrowLine(60,45)(110,70)
\Photon(60,45)(60,10){3}{5}
\Vertex(60,10){3}
\PhotonArc(60,45)(20,28,153){3}{6}
\Photon(91,60)(110,40){-3}{5}
\Text(22,72)[]{$p_1$}
\Text(98,72)[]{$p_2$}
\Text(52,25)[]{$q$}
\Text(100,37)[]{$k$}
\Text(128,37)[]{$+$}
\ArrowLine(130,70)(170,50)
\Line(170,50)(180,45)
\ArrowLine(190,50)(230,70)
\Line(180,45)(190,50)
\Photon(180,45)(180,10){3}{5}
\Vertex(180,10){3}
\PhotonArc(180,45)(25,28,153){3}{6}
\Photon(197,53)(216,33){-3}{5}
\Text(142,72)[]{$l$}
\Text(218,72)[]{$l$}
\Text(172,25)[]{$\gamma^*$}
\Text(223,37)[]{$\gamma$}
\Text(180,4)[]{$A,Z$}
\Text(248,37)[]{$+$}
\ArrowLine(250,70)(300,45)
\ArrowLine(300,45)(350,70)
\Photon(300,45)(300,10){3}{5}
\Vertex(300,10){3}
\PhotonArc(329,59)(14,28,205){3}{5}
\Photon(331,60)(350,40){-3}{5}
\Text(378,37)[]{$+\ \ldots$}
\end{picture}
\label{fig1}
\caption{Representatives of Feynman amplitudes with single virtual loops.}
\end{figure}

Representatives of the Feynman diagrams corresponding to the first type of corrections
are shown in Fig.~\ref{fig1}. This contribution was computed with help of the automatized
computer system SANC~\cite{Andonov:2004hi}. The system provided the set of form factors
calculated keeping the exact dependence on the lepton mass. The form factors are expressed
via a number of one--loop master integrals (Passarino--Veltman functions), which are
called from a SANC library. 
The infrared divergence in the relevant integrals is regularized 
by a fictitious photon mass $\lambda$. 
So, the virtual loop contribution takes the form
\ba
\frac{\dd\sigma^{\mathrm{Virt}}}{\dd\omega} &=& 
 \frac{Z^2\alpha^3}{2\pi}\int\limits_{-1}^{1}\dd c_1
\int\limits_{-1}^{1}\dd c_2\int\limits_{0}^{2\pi}\dd \varphi\,
\frac{|\vecc{p}_2|}{|\vecc{p}_1|}\frac{\omega}{Q^4}
    \frac{1}{16\pi^2e^4}\biggl( \FF^{(1)}_{\delta}(q_s,t_s,u_s)
- \FF^{(1)}_{11}(q_s,t_s,u_s)E_2^2
\nonumber \\
&-& \FF^{(1)}_{22}(q_s,t_s,u_s)E_1^2
+ \FF^{(1)}_{12}(q_s,t_s,u_s)E_2E_1
+ \FF^{(1)}_{21}(q_s,t_s,u_s)E_1E_2
  \biggr).
\ea
Here we adopt the SANC notation for the arguments and the normalization of 
the form factors. 
More details about the evaluation of form factors for such processes 
within SANC can be found in Ref.~\cite{Bardin:2005dp}.

\subsection{Soft Photon Contribution}
\label{subSect_S}

Using the phase space slicing method we define the soft photon contribution as the 
one of the process with emission of an additional photon with energy below
a certain value $\bar\omega$, which is small compared with the beam energy. 
In our calculations we perform the spicing in the laboratory
reference frame where the nucleus is at rest. Using the standard techniques
of soft photon emission calculations we get the corresponding correction,
which is factorized before the Born cross section:
\ba
\frac{\dd\sigma^{\mathrm{Soft}}}{\dd\omega} &=& 
\delta^{\mathrm{Soft}}\frac{\dd\sigma^{\mathrm{Born}}}{\dd\omega}, 
\qquad
\delta^{\mathrm{Soft}} = -\frac{\alpha}{4\pi^2}\biggl(I_{11} + I_{12} - 2I_{12}\biggr),
\nonumber \\
I_{11} &=& 4\pi\biggl[\ln\frac{2\bar\omega}{\lambda} 
+ \frac{1}{2\beta_1}\ln\biggl(\frac{1-\beta_1}{1+\beta_1}\biggr)\biggr],
\qquad
I_{22} = 4\pi\biggl[\ln\frac{2\bar\omega}{\lambda} 
+ \frac{1}{2\beta_2}\ln\biggl(\frac{1-\beta_2}{1+\beta_2}\biggr)\biggr],
\nonumber \\
I_{12} &=& \frac{2\pi}{1-m^2/(a_{12}p_1p_2)}
\biggl[2\ln\frac{2\bar\omega}{\lambda} \ln a_{12}
+ \frac{1}{4}\ln^2\biggl(\frac{1-\beta_1}{1+\beta_1}\biggr)
- \frac{1}{4}\ln^2\biggl(\frac{1-\beta_2}{1+\beta_2}\biggr)
\nonumber \\
&+&\Li{2}{1-\frac{a_{12}E_1}{v_{12}}(1+\beta_1)}
-\Li{2}{1-\frac{      E_2}{v_{12}}(1+\beta_2)}
\nonumber \\
&+&\Li{2}{1-\frac{a_{12}E_1}{v_{12}}(1-\beta_1)}
-\Li{2}{1-\frac{      E_2}{v_{12}}(1-\beta_2)}
\biggr],
\nonumber \\
\beta_{1,2} &=& \frac{|\vecc{p}_{1,2}|}{E_{1,2}}=\sqrt{1-\frac{m^2}{E_{1,2}}},\qquad
a_{12} = \frac{p_1p_2}{m^2}\biggl(1+\sqrt{1-\frac{m^4}{(p_1p_2)^2}}\biggr),
\nonumber \\
v_{12} &=& \frac{a_{12}p_1p_2-m^2}{a_{12}E_1-E_2}\, , \qquad
p_1p_2 = \frac{1}{2}Q^2 + \chi_2 - \chi_1 + m^2. 
\ea
The infrared divergence of the soft photon contribution is regularized by 
means of a fictitious photon mass $\lambda$, the same as in the virtual
loop contribution. One of the internal cross checks of the calculation is 
the cancellation of the dependence on this auxiliary parameter in the the
sum of two contributions.

\subsection{Double Bremsstrahlung Contribution}
\label{subSect_H}

\begin{figure}[htb]
\begin{picture}(400,80)(0,0)
\ArrowLine(10,70)(50,50)
\Line(50,50)(60,45)
\ArrowLine(70,50)(110,70)
\Line(60,45)(70,50)
\Photon(60,45)(60,10){3}{5}
\Vertex(60,10){3}
\Photon(37,56)(37,80){-3}{4}
\Photon(83,56)(83,80){3}{4}
\Text(47,73)[]{$k_1$}
\Text(74,73)[]{$k_2$}
\Text(128,37)[]{$+$}
\ArrowLine(130,70)(170,50)
\Line(170,50)(180,45)
\ArrowLine(190,50)(230,70)
\Line(180,45)(190,50)
\Photon(180,45)(180,10){3}{5}
\Vertex(180,10){3}
\Photon(157,56)(157,80){-3}{4}
\Photon(203,56)(203,80){3}{4}
\Text(167,73)[]{$k_2$}
\Text(195,73)[]{$k_1$}
\Text(248,37)[]{$+$}
\ArrowLine(250,70)(300,45)
\ArrowLine(300,45)(350,70)
\Photon(300,45)(300,10){3}{5}
\Vertex(300,10){3}
\Photon(315,52)(315,80){3}{4}
\Photon(331,60)(350,40){-3}{4}
\Text(344,37)[]{$k_2$}
\Text(308,73)[]{$k_1$}
\Text(378,37)[]{$+\ \ldots$}
\end{picture}
\label{fig2}
\caption{Representatives of Feynman amplitudes for double bremsstrahlung.}
\end{figure}
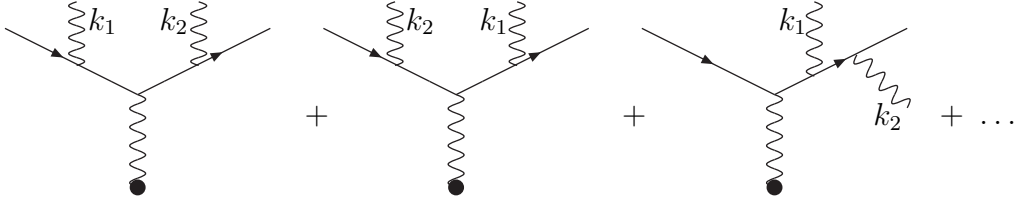

Here we start with the completely differential expression for the
matrix element squared. Some of the Feynman diagrams for this process 
are shown in Fig.~\ref{fig2}.
The two photons are treated in a symmetric way.
In particular, the condition $\omega_{1,2} > \bar\omega$ is applied for 
both the photons. The identity factor $1/2!$ is taken into account.
Cancellation of the dependence on the parameter $\bar\omega$
is checked numerically in the sum of the soft and hard contributions.
The contributions of double real photon emission is computed by means  
a Monte Carlo integrator based on the VEGAS algorithm~\cite{Lepage:1977sw}.
Seven-fold integration over the whole final state phase space (including integration
of the detected photon energy) is performed. The distribution in the detected photon
energy is extracted in course of the integration using weights provided by VEGAS for
each thrown kinematical point.  

\section{Numerical Results and Conclusions}
\label{Sect_Concl}

Summing up the considered above contributions we get the 1-loop corrected cross section in the
form
\ba
\frac{\dd\sigma^{\mathrm{Corr}}}{\dd\omega} = \frac{\dd\sigma^{\mathrm{Born}}}{\dd\omega}
+ \frac{\dd\sigma^{\mathrm{Virt}}}{\dd\omega} +\frac{\dd\sigma^{\mathrm{Soft}}}{\dd\omega}
+ \frac{\dd\sigma^{\mathrm{Hard}}}{\dd\omega}\, .
\ea
In Table~\ref{tab1} there are numerical results for the specific contributions
obtained for the following set of conditions:
\ba
\label{bare}
\begin{array}[b]{lcllcllcl}
E_1 &=& 190\GeV, & Z &=& 82, & Q^2_{\mathrm{max}} &=& 0.0075\GeV^2, \\ 
\bar{\omega} &=& 0.001\GeV,\quad & M_{\mathrm{max}} &=& 3.75\cdot m_\mu,\quad &
P^{\perp}_{\mathrm{min}} &=& 0.045\GeV, \\
m_l &=& m_\mu = 0.10566\GeV, & &
\end{array}
\ea 
where $M_{\mathrm{max}}$ is the maximal allowed invariant mass of the muon plus hard
photon final state system; $P^{\perp}_{\mathrm{min}}$ is the minimal 
allowed transverse momentum of the outgoing muon; $Z$ is the Pb nucleus charge. 
For the sake of simplicity, while computing the numbers for the Table  
we put a simple cut on the second hard photon energy: $\omega_2 < \omega_1$.
By subscripts $1$ and $2$ we denote the results obtained with $\bar\omega=10^{-3}$
and $\bar\omega=10^{-4}$, respectively. The relative corrections $\delta_{1,2}$ are
computed as
\ba
\delta_{1,2} = \frac{\dd\sigma^{\mathrm{Virt}}/\dd\omega
+ \dd\sigma_{1,2}^{\mathrm{Soft}}/\dd\omega
+ \dd\sigma_{1,2}^{\mathrm{Hard}}/\dd\omega}
{\dd\sigma^{\mathrm{Born}}/\dd\omega} \cdot 100\%.
\ea

\begin{table}[ht]
\begin{tabular}{|c|c|c|c|c|c||c|c|c|}
\hline
$\omega/E_1$ &
Born &
Virtual &
Soft$_1$ &
Hard$_1$ &
$\delta_1$, \% &
Soft$_2$ &
Hard$_2$ &
$\delta_2$, \% 
\\ \hline
0.3 & 15677(1)  & 76.8(4) & - 260.1(1) & 226.9(3) & $+$0.28 & 
-307.0(1) & 273.7(3) & $+$0.28 
\\ \hline
0.5 & 10836(1)  & 77.9(2) & - 319.0(1) & 280.0(3) & $+$0.36 & 
-377.4(1) & 338.1(3) & $+$0.36
\\ \hline
0.7 & 7337.7(1) & 76.9(2) & - 363.3(1) & 297.1(2) & $+$0.15 & 
-430.9(1) & 364.8(2) & $+$0.15 
\\ \hline
0.9 & 1267.4(1) & 20.5(1) & - 111.1(2) &  65.9(1) & $-$1.95 & 
-132.4(2) &  87.2(1) & $-$1.95 
\\ \hline
\end{tabular}
\label{tab1}
\caption{Contributions to the corrected differential cross section in pbarn/GeV
{\it versus} the photon energy fraction.}
\end{table}

For a realistic simulation of spacial resolution and cluster energy threshold
of the COMPASS calorimeter in addition to the conditions~(\ref{bare}),
we apply the following treatment of events with two hard photons:
\begin{itemize}

\item[1)]
 max$(\omega_1,\omega_2) \geq \omega_{\mathrm{th}}$,  
 {\it i.e.} at least one of the photons should have an energy exceeding the threshold; 

\item[2)] if both the photons have energies above the threshold and the
angle between their momenta is more than $\theta_{\gamma\gamma}$, the event
is dropped;

\item[3)] if the angle between their momenta is less than $\theta_{\gamma\gamma}$,
the reconstructed photon energy is taken as the sum of the two: $\omega=\omega_1+\omega_2$;

\item[4)] if one of the photon energies is below the threshold, the reconstructed photon energy is taken as the sum of the two: $\omega=\omega_1+\omega_2$.

\end{itemize}
The parameter values correspond to one of  data analysis procedures used by the
COMPASS experiment,
\ba\label{calo}
\omega_{\mathrm{th}} = 7\GeV, \qquad \theta_{\gamma\gamma} = 3~\mathrm{mrad}.
\ea

\begin{figure}[htb]
\begin{center}
\epsfig{file=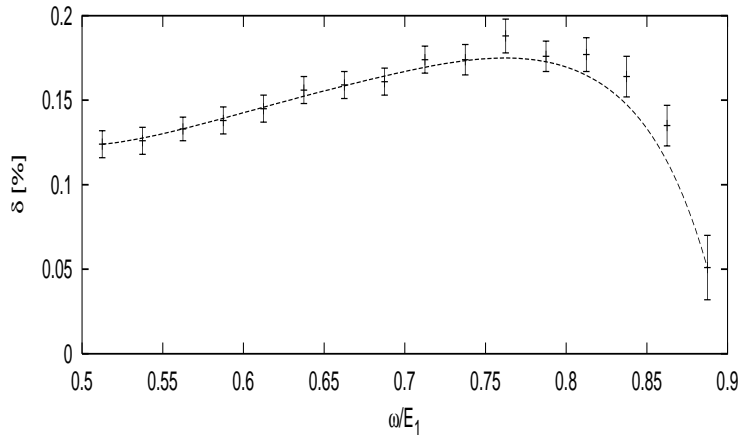,width=10cm,height=6cm}
\end{center}
\label{fig3}
\caption{Relative contribution of one-loop corrections for realistic set-up {\it vs.} the photon energy fraction.}
\end{figure}

For the realistic set--up, the size of the resulting correction is found to be below 
the one percent level. That is
due to the fact that the correction is proportional to $\alpha/(2\pi)$, and in our case there 
is no any enhancement factors. In particular, even so that the beam energy is so large 
compared with the muon mass, the contributions of the order $\order{\alpha\ln(E_1^2/m_\mu^2)}$
cancel out in the sum of different contributions due to destructive interference of the
initial and final state radiation. As can be seen from the Table~\ref{tab1} at the end
of the spectrum $(\omega\to E_1)$, where the phase space of additional hard photon 
emission is vanishing, we have a negative peak of the resulting radiation correction, 
which behaves there like $\alpha\ln((E_1-\omega)/E_1)$. 
But this peak is effectively washed out from
the end of the spectrum in Fig.~\ref{fig3}, because of the the additional conditions~(\ref{calo})
on event selections.

An analogous study was performed for the case of pion bremsstrahlung 
in Ref.~\cite{Akhundov:1984mr}, where a similar behavior and magnitude of the
one-loop corrections have been obtained within the scalar QED.

\subsection*{Acknowledgments}
I am grateful to the SANC team for providing codes for the form factors.
I would like to thank B.~Bardin, S.~Bondarenko, A.~Guskov, 
L.~Kalinovskaya, Z.~Kroumchtein, and A.~Olshevsky for fruitful discussions. 
This work was supported by the INTAS grant 03-51-4007
and by the RFBR grant 07-02-00932.

\end{document}